\documentclass[doublecol]{epl2} 

\title{Enhanced backscattering of a dilute Bose-Einstein condensate}
\shorttitle{Title} 

\author{Guillaume Labeyrie\inst{1}, Tomasz Karpiuk\inst{2,3}, Jean-Fran\c{c}ois Schaff\inst{1,4}, Beno\^it Gr\'emaud\inst{2,5,6}, Christian Miniatura\inst{1,2,5} \and Dominique Delande\inst{6}}
\shortauthor{G. Labeyrie \etal}

\institute{                    
  \inst{1} Institut Non Lin\'eaire de Nice, UMR 7335, UNS, CNRS; 1361 route des Lucioles, F-06560 Valbonne, France\\
  \inst{2} Center for Quantum Technologies, National University of Singapore, 3 Science Drive 2, Singapore 117543, Singapore\\
  \inst{3} Wydzia{\l} Fizyki, Uniwersytet w Bia{\l}ymstoku, ul. Lipowa 41, 15-424 Bia{\l}ystok, Poland\\
  \inst{4} Vienna Center for Quantum Science and Technology, Atominstitut, TU Wien, 1020 Vienna, Austria\\
  \inst{5} Department of Physics, National University of Singapore, 2 Science Drive 3, Singapore 117543, Singapore\\
  \inst{6} Laboratoire Kastler Brossel, Ecole Normale Sup\'erieure, CNRS, UPMC; 4 Place Jussieu, F-75005 Paris, France\\
}
\pacs{67.85.Hj}{Bose-Einstein condensates in optical potentials}
\pacs{42.25.Dd}{Wave propagation in random media}
\pacs{05.60.Gg}{Quantum transport}

\abstract{We study experimentally and numerically the quasi-bidimensional transport of a $^{87}$Rb Bose-Einstein condensate launched with a velocity $v_0$ inside a disordered optical potential created by a speckle pattern. A time-of-flight analysis reveals a pronounced enhanced density peak in the backscattering direction $-v_0$, a feature reminiscent of coherent backscattering. Detailed numerical simulations indicate however that other effects also contribute to this enhancement, including a ``backscattering echo'' due to the position-momentum correlations of the initial wave packet.}

\begin{document}

\maketitle

\section{Introduction}
Wave localization in non-homogeneous media is a general phenomenon, based on interference of multiple partial waves. While it is well understood in spatially periodic media, with e.g. band gap properties for electronic waves in solids and photons in photonic crystals, its generalization to disordered media has sparked a wide interest since the pioneering work of Anderson~\cite{Anderson1958} in 1958. Anderson localization has been investigated in a variety of systems including light~\cite{Storzer2006}, microwaves~\cite{Chabanov2000} or ultrasound~\cite{Hu2008}. Recently, dramatic progresses in the cooling and control of neutral atoms have allowed the investigation of the localization of quantum waves by disorder~\cite{Billy2008, Roati2008, Chabe2008, Kondov2011, Jend2012} (see~\cite{Shapiro2012} for a review). Observing Anderson localization in dimension 2 or 3 usually requires strong scattering, a condition difficult to realize experimentally. However, precursors of localization already exist in the regime of ``weak localization''~\cite{Meso}. Coherent backscattering (CBS), a constructive interference effect in the backscattering direction, is a hallmark of this regime and can be observed in a wide range of transport mean free path $\ell$. It is a signature of (quantum) coherent transport and consequently an important tool for characterizing scattering media, especially for studies of localization phenomena. CBS has been studied extensively with light~\cite{vanAlbada1985,Labeyrie1999}, ultrasound~\cite{Bayer1993}, seismic waves~\cite{Larose2004}, excitons~\cite{Langbein2002} and polaritons~\cite{Liew2009}.

\begin{figure}
\onefigure{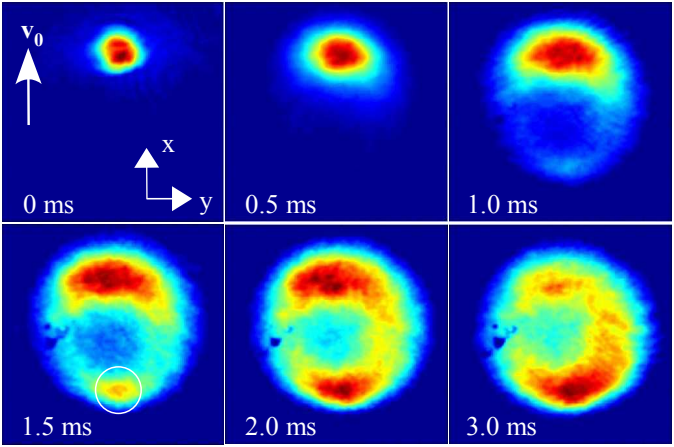}
\caption{(Color on line) Transport of a BEC in a 2D disorder. We measure the spatial atomic distribution after a 42 ms TOF, for various values of the time $\tau$ spent in the disordered potential. Each image, of field of view $1\times1$ mm, is the average of 30 shots with different disorder configurations. A peak in the backscattering direction is clearly observed for $t\geq1$ ms (as emphasized by the circle for $\tau = 1.5$ ms).}
\label{fig1}
\end{figure}

In this paper we report our experimental study of the transport of a Bose-Einstein condensate (BEC) in a bidimensional (2D) disorder, analyzed in momentum space through time-of-flight (TOF) absorption imaging. We observe an enhanced density peak in the backscattering direction (see Fig.~\ref{fig1}), a feature usually attributed to CBS. However, we show based on detailed numerical simulations that additional effects also lead to an enhanced scattering, rendering the direct observation of CBS in our regime a subtle issue. In particular, CBS can be a small effect in the usual observation channel (momentum space), but appear as a well localized bright spot in real space. 

\section{Experiment and simulations}

Our setup~\cite{Schaff2011} enables the production, in a Ioffe-Pritchard (IP) magnetic trap, of a BEC containing $10^5$ $^{87}$Rb atoms every 12 s. For the experiment described in this paper, a 2D optical dipole trap is turned on during the end of the evaporation phase. It is made of a vertical standing wave formed by crossing two laser beams at a small angle of $4.3^{\circ}$ in the vertical plane ($y$, $z$). The waists of the employed beams define an effective area of $2 \times 1.5$ mm$^2$ for the 2D waveguide. The laser wavelength is 766 nm, detuned by 14 nm to the blue of the D2 line of $^{87}$Rb. Once the BEC is loaded in the hybrid trap, we decompress the magnetic part by lowering the current in the IP trap. This produces a rather low density BEC, with a chemical potential $\mu = 270$ Hz for $N = 10^5$ atoms. We then release the BEC in the 2D waveguide by turning off the IP trap. Due to a shift of the magnetic trap center during the decompression, a velocity along $x$ is communicated to the BEC, which can be adjusted by selecting the shut-off time of the IP trap. We monitor the subsequent evolution of the BEC by absorption imaging. The images can be acquired immediately after the 2D trap shut-down to obtain the spatial atomic distribution, or after a 42 ms time-of-flight (TOF) to measure the velocity distribution. In the conditions of Fig.\ref{fig1}, the center-of-mass velocity of the initial wave packet is $v_0 =5.9$ mm/s (kinetic energy $E_k / h = 3.7$ kHz), and the velocity dispersions are $\Delta v_x = 2.6$ mm/s and $\Delta v_y = 1.4$ mm/s corresponding to the half-width of the Thomas-Fermi distribution. The spatial dimensions of the BEC are $\Delta x = 44 \mu$m and $\Delta y = 83 \mu$m. We study the BEC transport in presence of disorder by superimposing to the 2D waveguide a speckle pattern, formed by passing a vertical beam through a diffusive plate followed by a focusing lens. The disorder is here isotropic, contrary to the case of~\cite{RSV2010}. The speckle light is derived from the same laser as the 2D trap, frequency shifted by 160 MHz. The numerical aperture of the speckle optical system is 0.22, resulting in a disorder correlation length $\zeta = 0.54~\mu$m. In the plane of the BEC, the speckle intensity envelope is a Gaussian of waist = 0.74 mm. The total speckle power is 89 mW, yielding a speckle potential magnitude $V_{sp} = 2.7$ kHz. The speckle potential is turned on abruptly after the release of the BEC, for a variable duration $\tau$. Then, both speckle and 2D trap beams are switched off and the atomic distribution is measured after a 42 ms TOF.

In order to interpret the experimental results, we have performed numerical simulations of the full quantum dynamics of the atomic cloud. The speckle disorder is numerically generated as explained in~\cite{Kuhn:thesis}. The time-dependent Schr\"{o}dinger (if atom-atom interactions are neglected) or the Gross-Pitaevskii (when interactions are important, mostly just after the BEC is released from the trap) equation is numerically solved using a spatial discretization on a grid whose size is much smaller than both the atomic de Broglie wavelength and the disorder correlation length. Two methods have been used: direct integration of the resulting coupled time-dependent differential equations, or a split technique alternating phases in configuration space (where the spatial part of the Hamiltonian is diagonal) and momentum space (where the kinetic energy is diagonal). Both methods give the same results. For a given realization of the disorder, both the momentum distributions and the simulated spatial densities after the TOF display strong local fluctuations. This is why we perform a configuration averaging over few tens of disorder realizations. Since our simulation can either include or exclude the effect of interactions in the BEC, we can estimate their impact on CBS for the parameters used in the experiment. We find that for our parameters interactions are responsible for a small reduction of at most $10\%$ of the CBS contrast.

\begin{figure}
\onefigure{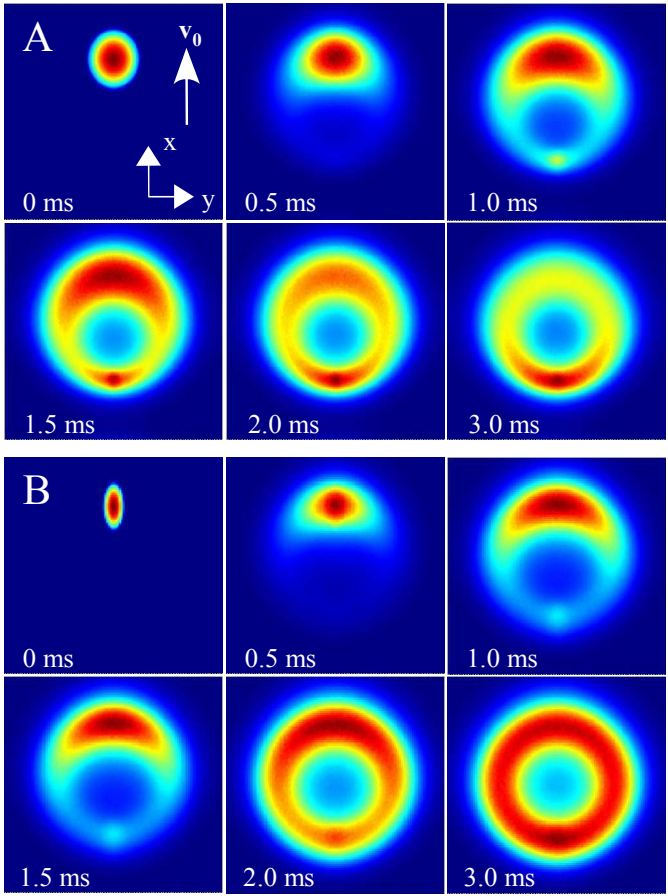}
\caption{(Color on line) Numerical results. (A) shows the spatial distribution after a 42 ms TOF (same field of view as in Fig.~\ref{fig1}). (B) is the true velocity distribution (infinite TOF). The parameters are those of the experiment (see text).}
\label{fig2}
\end{figure}

\section{Results} Fig.$~\ref{fig1}$ shows the measured distributions for increasing values of $\tau$ indicated in white. Each image actually corresponds to an average of 30 individual shots, between each the diffusive plate is rotated to achieve a different disorder configuration. The field of view is $1\times1$ mm. For each image the color scaling is adjusted such that it corresponds to the full dynamics of the signal. The white arrow on the first image indicates the direction of the initial center-of-mass velocity. Qualitatively, we observe that the spatial distribution after TOF rapidly develops into a ring as $\tau$ increases, which is consistent with the image of an elastic redistribution of the atomic velocities due to multiple scattering. As the diffusive component (the ring) develops, the initial wave packet centered at $v_0$ gets depleted. The ring radius is related to $v_0$ (we measure a radius of 259 $\mu$m, consistent with the mean atomic velocity) and its width to the velocity dispersion. In addition, the abrupt application of the speckle potential results in a ``disorder broadening'' of the ring, of the order of 30\% of its radius. Within 2 ms the diffusion ring has developed fully. However, the most striking observation is the clear appearance, after roughly 1 ms, of a \emph{peak in the backward direction} -$v_0$, the expected signature of the CBS effect~\cite{Cherroret2012}.

\begin{figure}
\onefigure{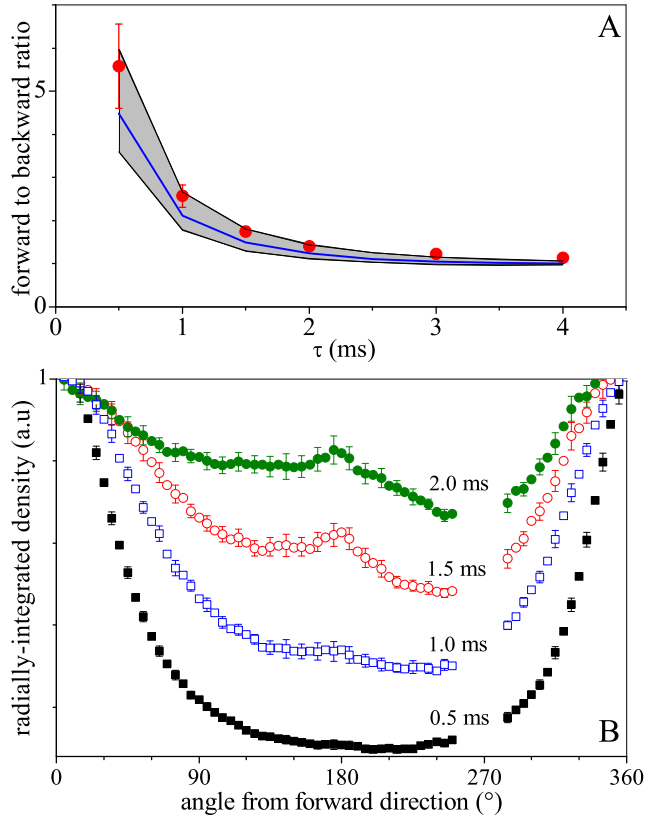} 
\caption{(Color on line) Transport dynamics. We compare in (A) the ratio of forward to backward scattering (see text), measured experimentally (circles) and numerically (line). The shaded area reflects the systematic uncertainty of $\pm 15\%$ on the speckle intensity. In B, we plot the angular distribution of the radially-integrated density (experimental data of Fig.~\ref{fig1}), for different values of $\tau$ (see text for details). The missing data around $270^{\circ}$ correspond to a saturated area on the images. The enhanced backscattering peak is clearly visible for $\tau \geq 1$ ms.}
\label{fig3}
\end{figure}

Fig.~\ref{fig2} shows the numerical results obtained using the experimental parameters. The upper panel corresponds to the spatial distribution after a 42 ms TOF, and the lower one to the true velocity distribution (infinite TOF). The spatial and color scales are the same as in Fig.\ref{fig1}. As can be seen, the simulation is very similar to the experiment. In particular, the dynamics of the population of the diffusion circle is well reproduced (see below and Fig.~\ref{fig3}A). Both finite and infinite TOF simulations yield an enhanced peak in the backscattering direction, but the peak at finite TOF is significantly higher and narrower. We will analyze the origin of this phenomenon in the \textit{Discussion} section.          

Fig.~\ref{fig3}A illustrates the dynamics of transport, as measured by the ratio of forward to backward scattering. This quantity is simply the ratio of the integrated densities of the ``top half'' of the images of Fig.~\ref{fig1} to the ``bottom half'', as defined by the horizontal line passing through the center of the diffusion circle. At times larger than the Boltzmann time~\cite{Cherroret2012}, one expects this ratio to approach unity. The circles correspond to the experimental data, while the line is from the numerics using the experimental parameters. The shaded area reflects the experimental uncertainty of $\pm 15\%$ on the speckle intensity. The quantitative agreement is good. We can extract from the simulations various quantities characterizing the transport, such as the momentum-averaged transport mean-free path $\ell = 7.0~\mu$m and transport time $\tau_{\mathrm{tr}} = 1.2~$ms. We then perform a radial integration of the experimental images of Fig.~\ref{fig1} to obtain the angular distributions reported in Fig.~\ref{fig3}B. For clarity, all curves have been scaled to 1 in the forward direction ($0^{\circ}$). The missing data points around $270^{\circ}$ correspond to a slightly saturated zone on the images, occuring at a fixed position and distorting the data in a limited angular range which we thus excluded. The amplitude of the error bars shown on the figure corresponds to $\pm 2$ standard deviations of the statistical fluctuations. As can be seen, an enhanced backscattering peak is significantly observed for $\tau \geq 1$ ms. The following section is devoted to the analysis of the mechanisms responsible for this enhancement.

\begin{figure}
\onefigure{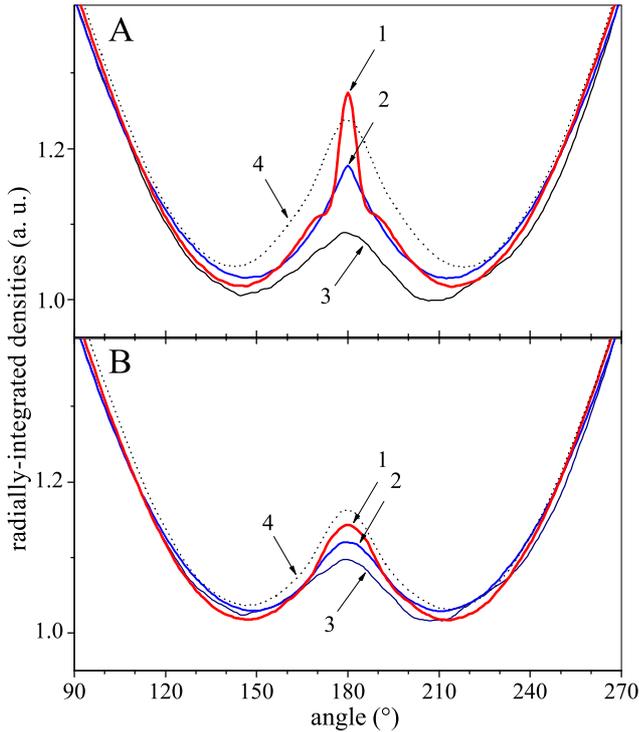} 
\caption{(Color on line) Contributions to enhanced backscattering (numerics). We compare the angular dependency around backscattering of the radially-integrated atomic density in four different situations: (1) quantum simulation using an initial wave packet with the same average velocity $v_0=$ 5.9 mm/s than in the experiment, isotropic spatial $\Delta x=\Delta y=54~\mu$m and momentum dispersion $\Delta v = 1.3~$mm/s and linear position-momentum correlation. The propagation time in the speckle is 1.5 ms; (2) classical simulation with the same initial conditions as in (1); (3) same as (2) but without position-momentum correlation; (4) quantum simulation using the initial state of the experiment, as in Fig.\ref{fig2}. (A) corresponds to a finite TOF = $t_{\mathrm{exp}}$, and (B) to the momentum distributions (infinite TOF).}
\label{fig4}
\end{figure}

\section{Discussion} Most of the behaviour observed in Fig.~\ref{fig2} is due to the specific properties of the initial wave packet created when releasing the BEC from the trap. Indeed, as shown in~\cite{CastinDum}, the wave packet roughly keeps the ``inverted parabola'' shape of the BEC in the Thomas-Fermi regime, but, more importantly, is characterized by a strong position-momentum correlation: the fastest atoms are located near the edges of the wave packet while the slowest ones lie at the center (in the frame co-moving with the center of mass of the wave packet). If one neglects the small quantum fluctuations (the wave packet is far from being minimal), there is even an almost perfect linear relation between the local velocity $v$ and the distance $d$ from the wave packet center. In other words, the initial wave packet of spatial size $\Delta x$ and velocity dispersion $\Delta v$ looks as if originating from a point source after a free expansion time $t_{\mathrm{exp}} = \Delta x/\Delta v$. This linear correlation has two important consequences on the backscattered atomic distribution after time-of-flight.

First, it yields a transient enhancement of the atomic density in the backscattering direction for TOFs around $t_{\mathrm{exp}}$. Indeed, if the speckle is applied during a short time $\tau$ such that the atomic displacement inside the speckle is negligible compared to $\Delta x$, all the particles which are backscattered will return to the origin (in the moving frame) after a time of flight $t_{\mathrm{exp}}$. Despite the initial velocity spread, all atoms will reassemble at the same point because of the quadratic dispersion relation of matter waves. We nickname this effect, reminiscent of the spin echo, the backscattering echo (BSE). Contrary to CBS, it is an \textit{incoherent} effect, maximum when the time of flight duration is equal to $t_{\mathrm{exp}}$. In the experiment, this time is $17$ ms along $x$ and $59$ ms along $y$. The 42 ms duration of the TOF lies between these values, so BSE is expected to be present as well as CBS in our situation.   

Second, the CBS contrast is also expected to be maximal for a TOF equal to $t_{\mathrm{exp}}$, in analogy with experiments in optics where the scattering sample is illuminated with a diverging light beam~\cite{Gross2007}. In this case, the CBS peak is observed with maximal contrast at the position of the effective point source. The CBS peak observed in the ``far field'' (in our case in momentum space) is the convolution of the peak at $t_{\mathrm{exp}}$ (of angular width $\simeq (k \ell)^{-1})$ by the angular spread of the atomic source $\Delta v/v_0$. To illustrate these effects, we plot in Fig.~\ref{fig4} the results of numerical simulations corresponding to different conditions: (1) a quantum simulation using a correlated initial wave packet, with an \textit{isotropic} velocity dispersion $\Delta v = 1.3$ mm/s; (2) a classical simulation using the same correlations and velocity spread as in (1); (3) same as (2) but with \textit{uncorrelated} position and momentum in the initial atomic distribution; (4) the quantum simulation of Fig.\ref{fig2} using the experimental (anisotropic) momentum distribution. We compare the angular distributions of the radially-integrated atomic densities after a TOF = $t_{\mathrm{exp}}$ (Fig.~\ref{fig3}A) and at infinite TOF (momentum distribution, Fig.~\ref{fig3}B). Simulations (1) and (4) are expected to include all effects, both classical and interferential. Simulation (1) yields for TOF = $t_{\mathrm{exp}}$ a narrow peak due to CBS while the momentum distribution exhibits only a broad feature, which is consistent with the convolution argument given above. This is a new situation where CBS is not visible in the usual observation channel (momentum space), but dramatic in real space where it appears as a localized bright spot. The characteristic CBS sharp feature of curve (1) in Fig.\ref{fig3}A is broadened in curve (4), due to the ``astigmatism'' resulting from the anisotropic momentum distribution of the initial wave packet. The classical simulation with initial correlations (2) yields a triangular peak at finite TOF, which is attributed to BSE. Indeed, this feature disappears when the correlations of the initial state are removed (3). A surprising observation is the residual broad peak in the backscattering direction that occurs in the classical case without correlations (3). Its width is found to be rather independent of the TOF, such that it is observed also in the momentum distribution (where BSE is absent). We attribute this enhancement to a mechanism similar to the ``opposition effect''~\cite{Gehrels1956}, also known as ``opposition surge''~\cite{NASA} or ``shadow hiding effect'' -- it is responsible for the increased brightness of the moon when it is full~\cite{Buratti1996} -- linked to the channelling of atoms inside the valleys of the speckle potential. It is a purely classical phenomenon which exists only when the average optical potential is comparable
to the initial kinetic energy, and only for relatively short times of the order of few $\tau_{\mathrm{tr}}$. We are currently pursuing a more detailed numerical and theoretical investigation of this unexpected effect.

Let us summarize the rather complicated situation described above: in the case of a large \textit{isotropic} velocity dispersion $\Delta v/v_0$, the narrow CBS peak may be observed by choosing a TOF = $t_{\mathrm{exp}}$. BSE, responsible for a non-interferential backscattering enhancement, may be suppressed by using an initial state without position-momentum correlations (such as e.g a thermal cloud). In our case of an anisotropic distribution, $t_{\mathrm{exp}}$ is not well defined and a sharp CBS peak can not be observed. A definite separation from the classical enhancement mechanisms presented above is then delicate.

\section{Conclusion} In conclusion, we presented in this paper our experimental and numerical findings on the quasi-2D transport of a BEC in a speckle optical potential, analyzed through time-of-flight. We observe a rapid diffusive behavior characterized by a transport time of the order of 1 ms, accompanied by the apparition of an enhanced backscattering peak. These observations are nicely reproduced by our numerical simulations using only experimental parameters. These simulations also indicate that the observed enhancement of backscattering is both due to CBS and two additional classical effects, one of which is a transient effect arising from the position-momentum correlations in the initial atomic wave packet~\cite{Jendrzejewski2012}. Further improvements in the experiment should allow the quantitative study of the dynamics of CBS and the determination of the coherence length, a key parameter in localization experiments. The present work deals with a rather dilute BEC, but the chemical potential can be increased by more than one order of magnitude to study the interplay of weak localization and non linearities. Further studies could include the search for the ``coherent forward scattering'' effect predicted at the onset of the Anderson localization regime~\cite{Karpiuk2012}.

\acknowledgments
We thank P. Vignolo and C. A. M\"{u}ller for fruitful discussions. This work was supported by CNRS and Universit\'e de Nice-Sophia Antipolis. We also acknowledge financial support from R\'egion PACA, F\'ed\'eration Wolfgang Doeblin. Computing resources have been provided by GENCI, IDRIS and IFRAF. C.M and B.G acknowledge funding from the CQT-CNRS LIA FSQL. CQT is a Research Center of Excellence funded by the Ministry of Education and the National Research Foundation of Singapore.

\end{document}